\documentclass{iopconfser}

\usepackage{graphicx}
\usepackage{amsmath,amssymb}
\usepackage{booktabs}
\usepackage{bm}

\begin{document}

\title{Configuration-Interaction Studies of Alpha Clustering in Atomic Nuclei}

\author{Brandon Roldan Tomei and Alexander Volya}

\affil{Department of Physics, Florida State University, Tallahassee, FL 32306, USA}

\email{bsr24@fsu.edu, avolya@fsu.edu}

\begin{abstract}
Clustering in atomic nuclei, and $\alpha$ clustering in particular, is a striking form of
self-organization in quantum many-body systems, emerging from nucleon--nucleon
interactions through the interplay of correlations, symmetry constraints, and continuum
coupling. We summarize a configuration-interaction approach in which cluster degrees of
freedom are built from center-of-mass boosted fragment wave functions and treated with
resonating-group techniques. The norm kernel plays an important role: Pauli blocking
suppresses cluster channels at small separations, restructures the relative-motion wave
functions, and governs the normalization of clustering strength. As the central
development we present a Green's function (GF) method that aids in resolving a long-standing
difficulty of harmonic-oscillator based configuration-interaction calculations that stems
from wave functions with Gaussian asymptotics. The method restores the correct
asymptotic behavior and extracts the asymptotic normalization coefficient (ANC) with
high accuracy even in very modest harmonic-oscillator spaces.
The value of the ANC is very sensitive to the exact position of the state relative to
threshold, and yet the GF method establishes the ANC as a structural observable that can
be determined precisely from the interior wave function using the experimentally or
otherwise known exact energy.
A self-consistent iteration, formulated as an integral-operator eigenvalue problem, is
also shown to recover both the energy and the ANC in cases when the energy of the state
is to be determined.
The method is illustrated for near-threshold states, such as
the weakly bound $1^-$ state of $^{16}$O, as well as for the ground states of $^{16}$O
and $^{20}$Ne.
\end{abstract}

\section{Introduction}
\begin{figure}[htb]
    \centering
    \includegraphics[width=0.58\textwidth]{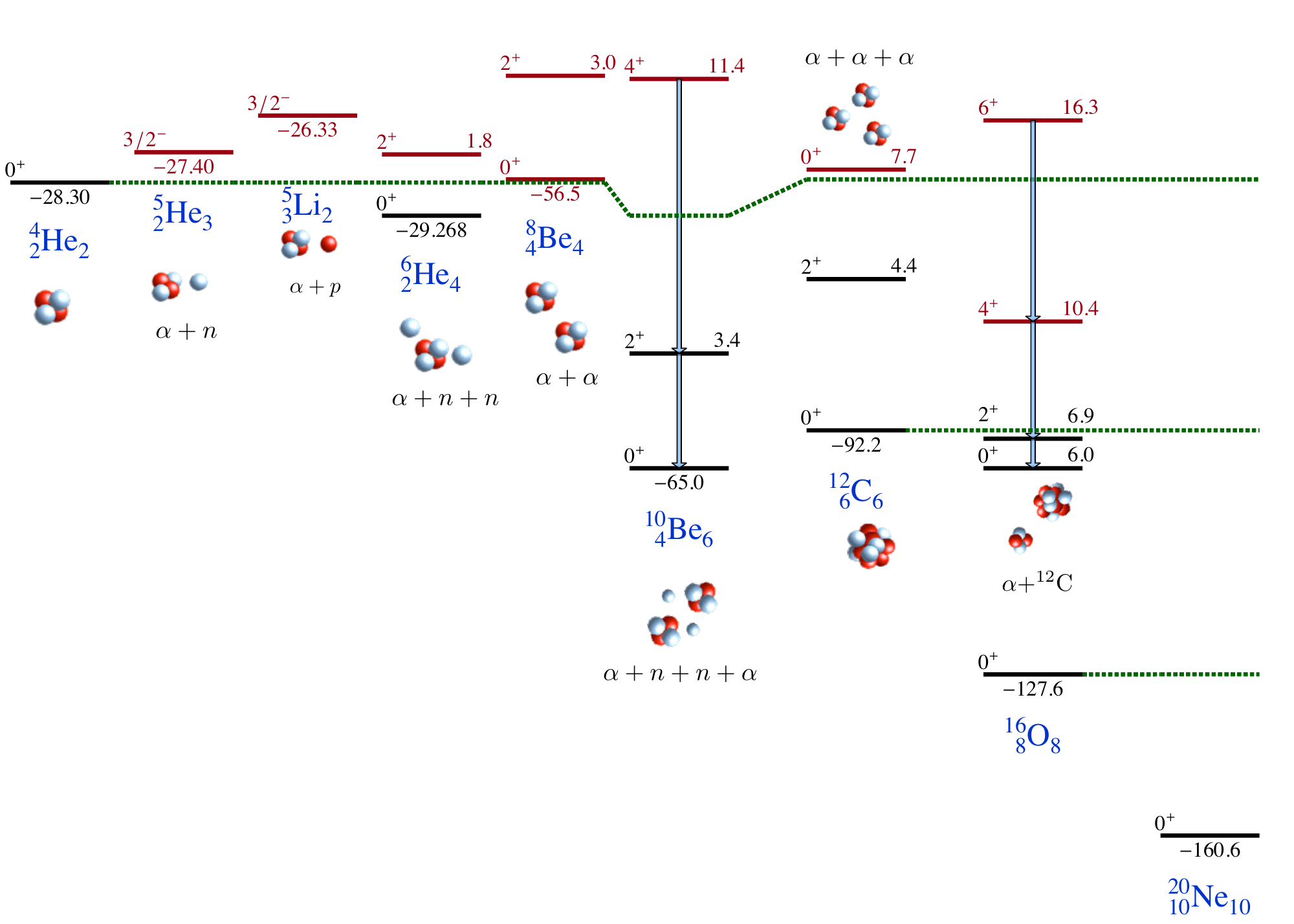}
    \caption{\label{fig:ikeda}Ikeda-style diagram illustrating clustering phenomena in
    light nuclei, with cluster-decay thresholds shown as dashed lines and decaying states
    highlighted in red. Spin-parity is shown on the left of each state and excitation
    energy on the right; total binding energy is indicated for ground states. All
    energies are in MeV. Adapted from Ref.~\cite{Otsuka2026}.}
\end{figure}

One of the enduring challenges of nuclear many-body physics is to elucidate how
structures, such as clusters, collective excitations, and shapes, emerge from the
underlying interactions among nucleons. Nuclear clustering, most notably the emergence of
$\alpha$-like correlations, has accompanied nuclear science since its early days. Many
molecular-like states in light nuclei appear near their respective cluster-decay
thresholds, a pattern succinctly summarized by the Ikeda diagram~\cite{Ikeda1968}; a
version highlighting thresholds and the corresponding cluster structures is shown in
Fig.~\ref{fig:ikeda}. Comprehensive discussions and historical references can be found in
the review~\cite{Freer2018}; the configuration-interaction perspective adopted here is
reviewed in Ref.~\cite{Otsuka2026}.

Microscopic calculations, starting with the Green's Function Monte Carlo demonstration of
$\alpha+\alpha$ structure in $^8$Be~\cite{Wiringa2000}, have shown that clustering emerges
directly from realistic nucleon--nucleon interactions; the no-core shell
model~\cite{Barrett2013}, symmetry-adapted frameworks~\cite{Launey2021}, lattice effective
field theory~\cite{Epelbaum2012}, and microscopic cluster methods rooted in the Resonating
Group Method (RGM)~\cite{Wheeler1937,Wildermuth1977} approach the problem from
complementary directions. Clustering also connects nuclear structure with reactions and
must be viewed from the open-quantum-systems
perspective~\cite{Johnson2020}: cluster states concentrate near decay thresholds, and
continuum coupling can realign wave functions toward clustering, a mechanism related to
superradiance~\cite{Auerbach2011} observed experimentally in isospin mirror
partners~\cite{Volya2022}. Resonance scattering with rare-isotope beams in inverse
kinematics now enables systematic studies of the clustering strength distribution in
nuclei on and away from the $4N$ line~\cite{Volya2022,Goldberg2022,Nurmukhanbetova2024}.
The near-threshold character of clustering makes these questions of paramount importance
for astrophysics, as exemplified by the $^{12}$C($\alpha,\gamma$)$^{16}$O reaction, which
determines the carbon-to-oxygen ratio in the universe~\cite{deBoer2017}.

In this contribution we briefly review how cluster degrees of freedom are incorporated
into the configuration-interaction (CI) framework, discuss the norm kernel and Pauli
blocking, and present, as the main new result, a Green's function method for
extracting asymptotic normalization coefficients (ANCs) from harmonic-oscillator based
CI calculations, in the context of a recent experimental determination of the $\alpha$
ANCs of the $^{16}$O and $^{20}$Ne ground states~\cite{Harris2025}.

\section{Cluster configurations in the configuration-interaction approach}
\label{sec:ci}
As in the standard shell model and no-core shell model, the single-particle harmonic
oscillator (HO) basis underlies the CI framework, and many-body states are superpositions
of Slater determinants. The symmetry structure of the many-body HO Hamiltonian allows an
exact factorization of the center-of-mass (CM) degree of freedom. While traditional
approaches retain only states whose CM is in the lowest oscillator mode,
$\Psi=\phi_{000}({\bf R})\,\Psi'$, clustering studies require more general
\emph{CM-boosted} configurations~\cite{Kravvaris2017,Kravvaris2017b}
\begin{equation}
\Psi_{n\ell m} = \phi_{n\ell m}({\bf R})\,\Psi',
\label{eq:boost}
\end{equation}
where $\phi_{n\ell m}$ is an HO wave function of the CM coordinate ${\bf R}$ carrying
$N_{\rm CM}=2n+\ell$ quanta and $\Psi'$ is the translationally invariant intrinsic state.
These states are constructed directly from existing shell model or no-core shell model
solutions using CM creation operators, with no additional diagonalization. When
the $\alpha$ particle is restricted to the $s^4$ configuration this construction reduces
to the analytic SU(3)-based cluster expansions widely used in the
literature~\cite{Smirnov1977,Volya2015}, but it is more general: clusters of any size and
with realistic internal structure can be boosted.

A two-cluster reaction channel with fragments $A_1$ and $A_2$ is expanded in channel basis
states built from boosted fragments,
\begin{equation}
\Phi_{n\ell m}
= \mathcal{A}\Bigl\{
\phi_{000}({\bf R})\,\phi_{n\ell m}(\bm{r})\,\Psi'^{(1)}\,\Psi'^{(2)}
\Bigr\},
\label{eq:channel}
\end{equation}
where $\bm{r}$ is the relative coordinate of the two fragment centers of mass, the
overall CM is in the lowest oscillator mode assuring non-spurious states, and
$\mathcal{A}$ antisymmetrizes over all nucleons. The recoupling of the fragment CM
coordinates into ${\bf R}$ and $\bm{r}$ is accomplished analytically with
Talmi--Moshinsky--Smirnov coefficients, so that Eq.~\eqref{eq:channel}, although written
in cluster form, remains a full many-body state expressed through Slater determinants.
The physical cluster channel is a superposition
$|\Xi_{\ell}\rangle=\sum_n \chi_n |\Phi_{n\ell}\rangle$, where the amplitudes $\chi_n$
follow from the RGM generalized eigenvalue problem~\cite{Wheeler1937,Wildermuth1977}
\begin{equation}
 \sum_{n'}\mathcal{H}_{nn'}\, \chi_{n'}  = E\sum_{n'}\mathcal{N}_{nn'}\,\chi_{n'},
 \qquad
 \mathcal{H}_{nn'} = \langle \Phi_{n\ell} | H | \Phi_{n'\ell} \rangle,
 \quad
 \mathcal{N}_{nn'} = \langle \Phi_{n\ell} |  \Phi_{n'\ell} \rangle,
 \label{eq:kernels}
\end{equation}
involving the Hamiltonian kernel $\mathcal{H}$ and the norm kernel $\mathcal{N}$. The
degree of clustering of a shell-model state $\Psi^{(A)}$ is quantified by the
spectroscopic factor (SF)
\begin{equation}
S_{\ell} = \bigl|\langle \Psi^{(A)} \,|\, \Xi_{\ell} \rangle\bigr|^{2}.
\label{eq:sf}
\end{equation}
This cluster-nucleon configuration-interaction strategy has been applied broadly: to
systematics of ground-state $\alpha$ SFs of $sd$-shell nuclei~\cite{Volya2015,
Kravvaris2017b}; to $\alpha$-cluster resonances in $^{19}$F and $^{20}$Ne, where
cross-shell interactions explain the distribution of clustering strength and the role of
extra nucleons~\cite{Goldberg2022,Volya2022}; and, in the
\textit{ab initio} context, to $\alpha+\alpha$ structure, scattering, and decay widths in
$^8$Be and to the triple-$\alpha$ structure of the Hoyle state in
$^{12}$C~\cite{Kravvaris2017,Kravvaris2019}. In what follows we concentrate on two
interrelated aspects at the heart of this program: the norm kernel and the asymptotic
normalization of cluster channels.

\section{Pauli blocking and the norm kernel}
\label{sec:norm}
\begin{figure}[htb]
\centering
\begin{minipage}[c]{0.42\textwidth}
\includegraphics[width=\linewidth]{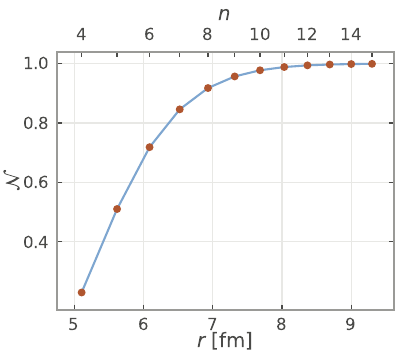}
\end{minipage}\hfill
\begin{minipage}[c]{0.505\textwidth}
\includegraphics[width=\linewidth]{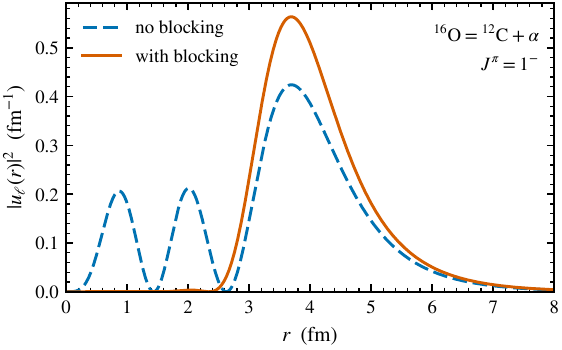}
\end{minipage}
\caption{\label{fig:norm}Impact of the Pauli principle on cluster
channels~\cite{Liguori}. Left: illustration of the norm-kernel behavior, based on the analytic results of
Refs.~\cite{Filippov2003,Filippov2004,Lashko2019}, for the $\ell=0$
$^{16}$O$+\alpha$ channel as a function of the separation distance, taken here as
$r=b\sqrt{2n+\ell+3/2}$; the top axis shows the corresponding radial quantum number
$n$. Pauli exchange suppresses $\mathcal{N}$ at small separations,
while $\mathcal{N}\to 1$ asymptotically. Right: relative-motion density
$|u_{\ell}(r)|^2$ in the $J^\pi=1^-$ $^{12}$C$+\alpha$ channel of $^{16}$O with and
without Pauli blocking.}
\end{figure}

The channel basis states \eqref{eq:channel} are not orthonormal: antisymmetrization among
all nucleons makes the norm kernel $\mathcal{N}$ in Eq.~\eqref{eq:kernels} non-trivial,
and its eigenvalues quantify how strongly a given cluster configuration is affected by the Pauli
principle. These eigenvalues can be obtained analytically by expanding the cluster
states in a symmetry-adapted basis, although the decomposition becomes increasingly involved for
more complex systems~\cite{Filippov2003,Filippov2004,Lashko2019}. Figure~\ref{fig:norm} (left) shows them
for the $\ell=0$ $^{16}$O$+\alpha$ channel, where the Pauli principle removes entirely
all states with fewer than eight oscillator quanta and strongly suppresses the norm at
small cluster--core separations, with eigenvalues approaching unity asymptotically.  
Importantly, harmonic-oscillator wave functions with increasing oscillator quantum number
asymptotically probe progressively larger distances, providing a natural correspondence
between the oscillator quantum number and the intercluster separation,
$r \simeq b\sqrt{2n+\ell+3/2}$, where $b$ is the oscillator length.
The
consequences for the channel wave function are illustrated in the right panel for the
$^{12}$C$+\alpha$ channel in $^{16}$O~\cite{Liguori}: the blocking suppresses the
interior oscillations of the relative-motion density and concentrates the clustering
strength at the nuclear surface.

The norm kernel is equally important for the interpretation of spectroscopic strength. In
restricted valence spaces the normalization
$\langle\Phi_{n\ell}|\Phi_{n\ell}\rangle$ of the relevant channel component is small, and
SFs evaluated with unrenormalized channels underestimate experimental data by an order of
magnitude; renormalizing the channel to unity, in the spirit of the orthogonality
condition model, restores both the magnitude and the characteristic trend of ground-state
$\alpha$ SFs across the $sd$ shell~\cite{Volya2015,Kravvaris2019}. Recent experimental
studies have started to probe the limits of this renormalization in situations where
Pauli blocking makes the channel normalization extremely
small~\cite{Nurmukhanbetova2024}. The interplay of interior suppression
and channel normalization directly influences the asymptotic properties of cluster
overlaps, and thus the decay widths and ANCs.

\section{Asymptotic normalization coefficients via the Green's function method}
\label{sec:gf}
The ANC quantifies the amplitude of the bound-state wave function in the asymptotic
region. For a bound state with separation energy defined by the wave number $\kappa$ and
Sommerfeld parameter $\eta$, the reduced radial wave function of the cluster relative
motion behaves as
\begin{equation}
u_{\ell}(r) \simeq C_{\ell}\, W_{-\eta,\,\ell+1/2}(2\kappa r), \qquad r\to\infty,
\label{eq:anc}
\end{equation}
where $W$ is the Whittaker function and $C_\ell$ is the ANC. ANCs are nearly
model-independent observables accessible in peripheral reactions and are key input for
astrophysical capture rates~\cite{Mukhamedzhanov2022,Brune1999,deBoer2017}. When
$u_{\ell}$ is the overlap function of a many-body state, $C_\ell$ is the physical ANC
probed in experiment; in a single-particle potential model the normalized relative-motion
wave function instead defines the single-particle ANC $\beta_\ell$; in the usual
single-particle factorization the two are related through the SF of
Eq.~\eqref{eq:sf}, $C_\ell = S_\ell^{1/2}\,\beta_\ell$. In the potential-model studies
below the extracted quantity is $\beta_\ell$; for convenience its numerical values are
quoted in the dimensionless convention
$\bar{\beta}_\ell \equiv 2^{-\eta}\,\beta_\ell/\sqrt{\kappa}$. Physical units are used in Sec.~\ref{sec:exp} for comparison with experiment.

Extracting ANCs from CI calculations faces a fundamental obstacle: the HO basis functions
fall off as Gaussians, so a variational HO-basis solution, even when its energy and
interior wave function are highly accurate~\cite{Payne1973}, never reproduces the
exponential-Coulomb asymptotics of Eq.~\eqref{eq:anc}. In practice the ratio
$u_{\ell}(r)/W_{-\eta,\ell+1/2}(2\kappa r)$ exhibits only a limited plateau before the
Gaussian collapse of the basis takes over, making the extraction inherently ambiguous,
especially for weakly bound states with extended tails.

The Green's function (GF) method resolves this problem. Green's function techniques
are a widely used strategy, and closely related ideas can be found in
Refs.~\cite{Timofeyuk2009,Timofeyuk2013,Burrows2024}. Here, the bound-state
Schr\"odinger equation is recast as an integral equation with the Coulomb Green's
function $g_{\ell}(r,r')$, built from the Whittaker functions,
\begin{equation}
u(r) = \frac{2\mu}{\hbar^2} \int_0^\infty dr'\, g_{\ell}(r, r')\, V(r')\, u(r'),
\qquad
g_{\ell}(r, r') = -\frac{\Gamma(\ell+1+\eta)}{2\kappa\,\Gamma(2\ell+2)}\,
M(2\kappa r_<)\, W(2\kappa r_>),
\label{eq:gf}
\end{equation}
where $\mu$ is the reduced mass and $V(r)$ is the remaining short-range nuclear
interaction. The indices
$({-\eta},\,{\ell+1/2})$ of the Whittaker functions are suppressed for brevity. Since the
correct asymptotics are encoded in $g_\ell$, taking $r\to\infty$ in Eq.~\eqref{eq:gf} and
comparing with Eq.~\eqref{eq:anc} yields a closed expression for the ANC as an interior
integral over the numerically known wave function; for the normalized potential-model
wave function this is the single-particle ANC,
\begin{equation}
\beta_{\ell} = -\frac{\mu}{\hbar^2 \kappa}\,
\frac{\Gamma(\ell+1+\eta)}{\Gamma(2\ell+2)}
\int_0^\infty dr'\, M_{-\eta, \ell+1/2}(2\kappa r')\, V(r')\, u(r').
\label{eq:ancgf}
\end{equation}
The integral is dominated by the nuclear interior, where the HO expansion is reliable,
making the incorrect asymptotic behavior of the input wave function largely irrelevant; 
Eq.~\eqref{eq:gf} can also be applied iteratively to correct the wave-function tail itself,
but we found no case in which such iteration was necessary.


\begin{figure}[htb]
    \centering
    \includegraphics[width=0.58\textwidth]{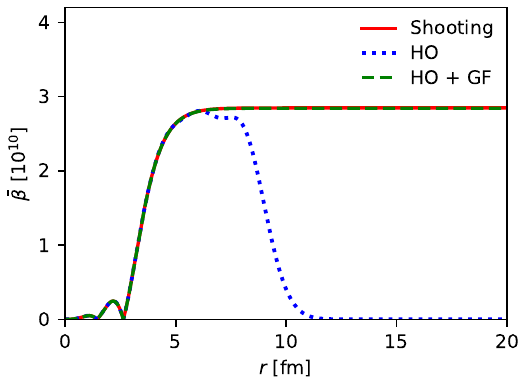}
    \caption{\label{fig:gf}The weakly bound $J^\pi=1^-$ state of $^{16}$O, only $\sim$45~keV below the $\alpha$-emission threshold~\cite{Tilley1993}, in the $^{12}$C$+\alpha$ potential model. Dimensionless single-particle ANC $\bar{\beta}$, obtained from the asymptotic ratio in Eq.~\eqref{eq:anc}, versus radius $r$; the GF-corrected result reproduces the reference across the full plateau where the bare HO extraction fails. ``Shooting'' is the exact solution obtained from the shooting method, ``HO'' is obtained using the harmonic-oscillator basis expansion, and ``HO + GF'' is the GF-corrected HO result obtained using Eq.~\eqref{eq:gf}.}
\end{figure}

Figure~\ref{fig:gf} illustrates the method for a demanding case: the $1^-$ state of
$^{16}$O bound by only $\sim$0.045~MeV relative to the $\alpha$
threshold~\cite{Tilley1993}, treated in a $^{12}$C$+\alpha$ Woods--Saxon potential model
with the number of interior nodes chosen as the minimum allowed by the Pauli principle.
All potential-model
calculations here use Woods--Saxon geometry $R=2.675$~fm ($^{12}$C$+\alpha$) or
$R=3.099$~fm ($^{16}$O$+\alpha$) with diffuseness $a=0.65$~fm, the Coulomb potential of a
uniformly charged sphere of the same radius, and reduced masses $\mu=2795.08$ and
$2981.33$~MeV, respectively. The depth is adjusted for each state to reproduce the
physical $\alpha$ separation energy with the minimum allowed number of nodes:
$(n,\ell)=(2,1)$ with $V_0=82.43$~MeV for the $^{16}$O $1^-$,
$(2,0)$ with $80.04$~MeV for the $^{16}$O g.s., $(4,0)$ with $134.48$~MeV for the
$^{20}$Ne g.s., and $(2,4)$ with $127.60$~MeV for the $^{20}$Ne $4^+$. The oscillator
length $b$ of the HO basis is optimized variationally for the target state,
independently at each basis size $N_{\rm basis}$, the number of radial oscillator
functions (e.g.\ $b=1.39$~fm for the $^{16}$O
$1^-$ at $N_{\rm basis}=7$). The exact
shooting-method solution serves as reference, and a Lagrange-mesh
calculation~\cite{Baye2015} provides an independent check. While the bare HO expansion
leaves only a narrow, state-dependent range of distances in which the ANC can be
determined, both the direct integral \eqref{eq:ancgf} and the GF-corrected wave function (``HO + GF'' in
Fig.~\ref{fig:gf}), obtained by applying Eq.~\eqref{eq:gf} to the HO solution at the
exact energy, reproduce the reference value $\bar{\beta}=2.848\times10^{10}$. 
Two aspects of the method deserve a closer look: the sensitivity of the
extraction to the input energy, which becomes extreme near threshold, and the
possibility of determining the energy self-consistently within the method itself. These
are addressed below.

\subsection{Energy sensitivity and universal near-threshold scaling}
\label{sec:energy}
The separation energy enters Eq.~\eqref{eq:ancgf} only through $\kappa$ and $\eta$, and
the exact energy impacts the value of the ANC very significantly. As discussed in
Ref.~\cite{Okolowicz2012}, splitting the norm of the bound state into interior and
exterior contributions, and using the $\eta\rightarrow\infty$ asymptotics of the
Whittaker function, the requirement that the exterior norm remains finite leads to the
universal near-threshold behavior of the single-particle ANC,
\begin{equation}
\beta_{\ell}(\eta) = \sqrt{\eta}\; e^{\,\eta\ln\eta-\eta}
\sum_{i=0}^{\infty} \frac{a_i(n,\ell)}{\eta^{i}},
\label{eq:threshold_expansion}
\end{equation}
where a universal prefactor multiplies a Maclaurin series in $1/\eta$ whose coefficients
$a_i$ contain the structure information of the state. 

Near the threshold,
$\kappa\to0$ and $\eta\to\infty$, and the charged-particle ANC varies
by orders of magnitude for keV-scale changes of the state's energy; for a neutral particle
($\eta=0$) the dependence enters through $\kappa$ alone and is a much milder power law.
The interior wave function, in contrast, is essentially insensitive to the energy in a
narrow near-threshold region; the entire energy dependence of the extraction resides in the
Green's function kernel, where it is treated exactly. Consequently, when the exact
separation energy is supplied, Eq.~\eqref{eq:ancgf} recovers the ANC with excellent
precision even from wave functions too crude to determine the energy variationally. For
the $^{16}$O $1^-$ state at $N_{\rm basis}=7$, for example, the variational energy
$E_{\rm HO}=-0.021$~MeV misses the exact $E=-0.045$~MeV by only 24~keV, yet the
extraction at $E_{\rm HO}$ overshoots the ANC by seven orders of magnitude, following
Eq.~\eqref{eq:threshold_expansion}, while the same wave function evaluated at the exact
energy gives $\bar{\beta}$ to $0.07\%$. Thus, an accurate determination of the ANC
requires accurate knowledge of the energy; in practical
applications the experimental separation energy is typically known.
Table~\ref{tab:conv} summarizes the accuracy achieved in this way for four
representative states, from the barely bound $1^-$ of $^{16}$O to the well-bound
ground states of $^{16}$O and $^{20}$Ne: with the physical separation energy as input,
percent-level accuracy is reached already with $N_{\rm basis}\sim10$ radial oscillator
functions. For $^{20}$Ne the minimal number of quanta allowed by the Pauli principle
(Wildermuth rule) is 8, which explains the large errors seen for $N_{\rm basis}=5$ and 7.
It is important to stress that this approach works in a way similar to spectroscopic
factors and reduced transition rates, which are commonly used to separate structure from
reaction kinematics in a perturbative way. The GF method decouples the
energy-sensitive kinematics of tunneling, contained in the wave-function tail, from the
structure information, and thus allows important physics to be established even from
models that are not exact enough to determine the energy to the needed precision.

\begin{table}[htb]
\centering
\caption{\label{tab:conv}Percent deviation from the exact reference ANC of the direct GF
integral \eqref{eq:ancgf}, evaluated with HO-basis wave functions of increasing basis
size $N_{\rm basis}$ (number of radial oscillator functions) and the physical separation
energy as input, for four $\alpha+{\rm core}$ states; energies $E$ (in MeV) are relative to the $\alpha$ threshold.}
\smallskip
\begin{tabular}{ccccc}
\toprule
 & $^{16}$O $1^-$ & $^{20}$Ne $4^+$ & $^{16}$O g.s. & $^{20}$Ne g.s. \\
$N_{\rm basis}$ & ($E=-0.045$) & ($E=-0.48$) & ($E=-7.16$) & ($E=-4.73$) \\
\midrule
5  & 1.12\%   & 6.13\%  & 2.06\%   & 23.0\%  \\
7  & 0.068\%  & 1.27\%  & 0.188\%  & 5.69\%  \\
10 & 0.042\%  & 0.484\% & 0.066\%  & 1.32\%  \\
25 & $<$0.01\% & $<$0.01\% & $<$0.01\% & $<$0.01\% \\
\bottomrule
\end{tabular}
\end{table}

\subsection{Self-consistent iteration}
\label{sec:iter}
When the energy of the state is not constrained by experiment and is to be determined
from the Hamiltonian in a fully consistent process, the energy and the ANC can be obtained
self-consistently from Eq.~\eqref{eq:gf}, which states that the bound state is an
eigenvector, with eigenvalue exactly 1, of the integral operator $\hat{L}(E)$ with kernel
$(2\mu/\hbar^2)\,g_{\ell}(r,r')\,V(r')$; the unit eigenvalue is realized only at the
correct energy. However, direct iteration of $u = \hat{L}(E)u$ is unstable whenever
$\hat{L}$ contains eigenmodes with $|\lambda| > 1$, since these modes grow relative to
the physical state with $\lambda \simeq 1$. To identify the origin of this instability,
we diagonalized both $\hat{L}$ and the Hamiltonian in the same truncated HO space. For
the $^{16}$O $1^-$ state at $N_{\rm basis} = 7$, the kernel spectrum is
$\lambda = \{3.33,\, 1.67,\, 1.00,\, 0.58,\, \dots\}$; the two $\lambda > 1$ modes are
dominated by the more deeply bound Hamiltonian eigenstates, while the squared overlap
of the $\lambda \simeq 1$ eigenvector with the desired state is $0.999$. This mirrors
the coupling-strength eigenvalue analysis of Weinberg~\cite{Weinberg1963}: the number of
eigenvalues above unity equals the number of states lying below the target energy.
This motivates the practical procedure: the known lower-energy Hamiltonian eigenstates
are projected out after each application of $\hat{L}$, the state is renormalized, and the
energy is updated through $E = \langle u|H|u\rangle / \langle u|u\rangle$, which is fed
back into the kernel through $\kappa$ and $\eta$. 

Starting from a very small HO basis, in which the initial variational estimate even
places the state above threshold, ten iterations recover for the barely subthreshold
$^{16}$O $1^-$ state the energy to better than $0.02\%$ and the ANC to better than
$0.25\%$ of the exact reference.

\section{Alpha clustering in the ground states of $^{16}$O and $^{20}$Ne}
\label{sec:exp}
The techniques discussed above meet experiment in a recent benchmark study of
ground-state $\alpha$ clustering~\cite{Harris2025}, where the sub-Coulomb
$\alpha$-transfer reaction $^{12}$C($^{20}$Ne,$^{16}$O)$^{16}$O at 1~MeV/nucleon was used
to measure, in a nearly model-independent way~\cite{Brune1999,Mukhamedzhanov2022}, the
product of the squared $\alpha$ ANCs of the $^{16}$O and $^{20}$Ne ground states. The
experimental result, expressed through the geometric mean that retains the scale of an
individual ANC, is
\begin{equation}
\Bigl[\bigl(C_{\alpha,^{12}{\rm C}}^{^{16}{\rm O}}\bigr)^2
\bigl(C_{\alpha,^{16}{\rm O}}^{^{20}{\rm Ne}}\bigr)^2\Bigr]^{1/4}
= 1.75(7)(9)\times10^{3}~{\rm fm}^{-1/2},
\label{eq:exp}
\end{equation}
and it was benchmarked, in a blind analysis, against three microscopic approaches:
nuclear lattice effective field theory~\cite{Epelbaum2012}, the symmetry-adapted no-core
shell model~\cite{Launey2021}, and the cluster-nucleon configuration interaction of
Sec.~\ref{sec:ci}~\cite{Kravvaris2019}, which predicted geometric means of $1.20(25)$,
$0.89(14)$, and $1.53(13)\times10^{3}$~fm$^{-1/2}$, respectively; the large experimental
value provides direct evidence that both ground states are dominated by $\alpha$-cluster
correlations.

The GF method places the potential-model limit of this comparison on firm numerical
ground. Using Woods--Saxon $\alpha+{\rm core}$ potentials with depths adjusted to the
physical $\alpha$ separation energies and Pauli-consistent node numbers, the direct
integral \eqref{eq:ancgf} applied to HO-basis wave functions gives
$C_{\alpha,^{12}{\rm C}}^{^{16}{\rm O}}=556$~fm$^{-1/2}$ and
$C_{\alpha,^{16}{\rm O}}^{^{20}{\rm Ne}}=5497$~fm$^{-1/2}$, corresponding to the
geometric mean $1.749\times10^{3}$~fm$^{-1/2}$, in remarkable agreement with the
experimental value \eqref{eq:exp} and reinforcing the conclusion of
Ref.~\cite{Harris2025} that ground-state $\alpha$ clustering in these nuclei approaches
its maximal, fully preformed value. At the same time, the spread among microscopic
predictions, most pronounced for the compact $^{16}$O ground state, traces back to the
physics of Sec.~\ref{sec:norm}: the normalization and non-orthogonality of cluster
channels, Pauli suppression of the interior, and the sensitivity of the ANC
to ground-state structure and configuration mixing.

\section{Summary and outlook}
Configuration-interaction techniques with explicitly constructed cluster channels provide
a unified microscopic view of $\alpha$ clustering in light nuclei. The norm
kernel plays a central role: Pauli blocking suppresses cluster configurations at small
separations, reshapes relative-motion wave functions, and sets the normalization of
spectroscopic strength. We presented a Green's function method that removes the
long-standing difficulty in extracting asymptotic normalization coefficients from
harmonic-oscillator based calculations; the ANC follows from an interior integral
insensitive to the incorrect Gaussian tail of the basis, reaching excellent accuracy
in bases with essentially the minimal number of oscillator functions needed to describe
the relative motion. We demonstrated this in a number of examples for states with
energies relative to threshold ranging from tens of keV to several MeV.

The ANC itself is very sensitive to the energy of the state relative to the decay
threshold; in the near-threshold region it follows a universal scaling with the
Sommerfeld parameter.
Thus an accurate energy is an essential ingredient of the ANC determination. The GF
method does not remove this sensitivity but handles it exactly: the tunneling kinematics
resides in the Green's function kernel, and the ANC emerges as a structural observable
determined by the interior wave function, much as energy-sensitive decay widths are
reduced to spectroscopic factors, or steeply falling cross sections to astrophysical
S-factors. Reliable ANCs of bound and subthreshold states are, in
turn, a key component of broad experimental strategies, such as comprehensive R-matrix
analyses of astrophysical reactions, where subthreshold resonances often control the
cross section at stellar energies~\cite{deBoer2017}.

Work is in progress to combine the GF method with
fully microscopic cluster-nucleon CI overlap functions; the combined progress of theory
and experiment~\cite{Harris2025} opens quantitative tests of clustering with implications
for nuclear structure, astrophysics, and open quantum many-body systems.

\section*{Acknowledgments}
\begin{sloppypar}
We thank our collaborators M.~Barbui, V.~Z.~Goldberg, K.~Kravvaris,
Yu.~A.~Lashko, K.~Liguori, A.~K.~Nurmukhanbetova, and G.~V.~Rogachev. This material is based upon work
supported by the U.S. Department of Energy, Office of Science, Office of Nuclear
Physics, under Award Number DE-SC0009883, and by the U.S. Department of Energy
under Award Number DE-SC0026615 (Genesis Mission).\footnote{This report was prepared as
an account of work sponsored by an agency of the United States Government. Neither the
United States Government nor any agency thereof, nor any of their employees, makes any
warranty, express or implied, or assumes any legal liability or responsibility for the
accuracy, completeness, or usefulness of any information, apparatus, product, or process
disclosed, or represents that its use would not infringe privately owned rights.
Reference herein to any specific commercial product, process, or service by trade name,
trademark, manufacturer, or otherwise does not necessarily constitute or imply its
endorsement, recommendation, or favoring by the United States Government or any agency
thereof. The views and opinions of authors expressed herein do not necessarily state or
reflect those of the United States Government or any agency thereof.}
\end{sloppypar}


\begin{thebibliography}{99}
\itemsep=0pt
\bibitem{Ikeda1968} Ikeda K, Takigawa N and Horiuchi H 1968 {\it Prog. Theor. Phys. Suppl.} {\bf E68} 464
\bibitem{Freer2018} Freer M, Horiuchi H, Kanada-En'yo Y, Lee D and Mei{\ss}ner U-G 2018 {\it Rev. Mod. Phys.} {\bf 90} 035004
\bibitem{Otsuka2026} Otsuka T, Volya A and Itagaki N 2026 Theoretical studies of $\alpha$ clustering in nuclei and beyond {\it Eur. Phys. J.} A {\bf 62} 108
\bibitem{Wiringa2000} Wiringa R B, Pieper S C, Carlson J and Pandharipande V R 2000 {\it Phys. Rev.} C {\bf 62} 014001
\bibitem{Barrett2013} Barrett B R, Navr\'atil P and Vary J P 2013 {\it Prog. Part. Nucl. Phys.} {\bf 69} 131
\bibitem{Launey2021} Launey K D, Mercenne A and Dytrych T 2021 {\it Annu. Rev. Nucl. Part. Sci.} {\bf 71} 253
\bibitem{Epelbaum2012} Epelbaum E, Krebs H, L\"ahde T A, Lee D and Mei{\ss}ner U-G 2012 {\it Phys. Rev. Lett.} {\bf 109} 252501
\bibitem{Wheeler1937} Wheeler J A 1937 {\it Phys. Rev.} {\bf 52} 1107
\bibitem{Wildermuth1977} Wildermuth K and Tang Y C 1977 {\it A Unified Theory of the Nucleus} (Braunschweig: Vieweg)
\bibitem{Johnson2020} Johnson C W {\it et al} 2020 {\it J. Phys. G: Nucl. Part. Phys.} {\bf 47} 123001
\bibitem{Auerbach2011} Auerbach N and Zelevinsky V 2011 {\it Rep. Prog. Phys.} {\bf 74} 106301
\bibitem{Volya2022} Volya A, Barbui M, Goldberg V Z and Rogachev G V 2022 {\it Commun. Phys.} {\bf 5} 322
\bibitem{Goldberg2022} Goldberg V Z {\it et al} 2022 {\it Phys. Rev.} C {\bf 105} 014615
\bibitem{Nurmukhanbetova2024} Nurmukhanbetova A K {\it et al} 2024 {\it EPJ Web Conf.} {\bf 311} 00022
\bibitem{deBoer2017} deBoer R J {\it et al} 2017 {\it Rev. Mod. Phys.} {\bf 89} 035007
\bibitem{Harris2025} Harris E {\it et al} 2025 Quantifying $\alpha$ clustering in the ground states of $^{16}$O and $^{20}$Ne {\it Preprint} arXiv:2507.17059 [nucl-ex]
\bibitem{Kravvaris2017} Kravvaris K and Volya A 2017 {\it Phys. Rev. Lett.} {\bf 119} 062501
\bibitem{Kravvaris2017b} Kravvaris K and Volya A 2017 {\it J. Phys.: Conf. Ser.} {\bf 863} 012016
\bibitem{Smirnov1977} Smirnov Yu F and Tchuvil'sky Yu M 1977 {\it Phys. Rev.} C {\bf 15} 84
\bibitem{Volya2015} Volya A and Tchuvil'sky Yu M 2015 {\it Phys. Rev.} C {\bf 91} 044319
\bibitem{Kravvaris2019} Kravvaris K and Volya A 2019 {\it Phys. Rev.} C {\bf 100} 034321
\bibitem{Filippov2003} Filippov G F, Lashko Yu A, Korennov S V and Kat\={o} K 2003 {\it Few-Body Syst.} {\bf 33} 173
\bibitem{Filippov2004} Filippov G and Lashko Y 2004 {\it Phys. Rev.} C {\bf 70} 064001
\bibitem{Lashko2019} Lashko Yu A, Vasilevsky V S and Filippov G F 2019 {\it Ann. Phys.} {\bf 409} 167930
\bibitem{Liguori}
Liguori K 2023 {\it Asymptotic Normalization Coefficients for $\alpha$ Clusters}
Honors thesis, Florida State University
\bibitem{Mukhamedzhanov2022} Mukhamedzhanov A M and Blokhintsev L D 2022 {\it Eur. Phys. J.} A {\bf 58} 29
\bibitem{Timofeyuk2009} Timofeyuk N K 2009 {\it Phys. Rev. Lett.} {\bf 103} 242501
\bibitem{Timofeyuk2013} Timofeyuk N K 2013 {\it Phys. Rev.} C {\bf 88} 044315
\bibitem{Burrows2024} Burrows M, Launey K D, Mercenne A, Baker R B, Sargsyan G H, Dytrych T and Langr D 2024 {\it Phys. Rev.} C {\bf 109} 014616
\bibitem{Brune1999} Brune C R, Geist W H, Kavanagh R W and Veal K D 1999 {\it Phys. Rev. Lett.} {\bf 83} 4025
\bibitem{Payne1973} Payne G L and Schlessinger L 1973 {\it J. Comput. Phys.} {\bf 13} 266
\bibitem{Baye2015} Baye D 2015 {\it Phys. Rep.} {\bf 565} 1
\bibitem{Tilley1993} Tilley D R, Weller H R and Cheves C M 1993 {\it Nucl. Phys.} A {\bf 564} 1
\bibitem{Okolowicz2012} Oko{\l}owicz J, Michel N, Nazarewicz W and P{\l}oszajczak M 2012 {\it Phys. Rev.} C {\bf 85} 064320
\bibitem{Weinberg1963} Weinberg S 1963 {\it Phys. Rev.} {\bf 131} 440
\end{thebibliography}
\end{document}